\documentstyle[preprint,aps]{revtex}

\def\lsim{\raise0.3ex\hbox{$\;<$\kern-0.75em\raise-1.1ex
\hbox{$\sim\;$}}}
\def\gsim{\raise0.3ex\hbox{$\;>$\kern-0.75em\raise-1.1ex
\hbox{$\sim\;$}}}

\begin{document}

\preprint{
\vbox{
\hbox{Fermilab-Pub-03/016-T} \vskip -0.2cm
\hbox{IFT-P.002/2003}\vskip -0.2cm
\hbox{hep-ph/0301210}\vskip -0.2cm
}}
\title{
The Complementarity of 
Eastern and Western Hemisphere
Long-Baseline Neutrino Oscillation Experiments
}

\author{
Hisakazu Minakata$^1$,
Hiroshi Nunokawa$^2$,
Stephen Parke$^3$
}

\address{
$^1$
\sl Department of Physics, Tokyo Metropolitan University \\[-0.3cm]
1-1 Minami-Osawa, Hachioji, Tokyo 192-0397, Japan\\[-0.3cm]
minakata@phys.metro-u.ac.jp
}

\address{
$^2$
\sl Instituto de F\'{\i}sica Te\'orica,
Universidade Estadual Paulista \\[-0.3cm]
Rua Pamplona 145, 01405-900 S\~ao Paulo, SP Brazil\\[-0.3cm]
nunokawa@ift.unesp.br
}

\address{
$^3$
\sl Theoretical Physics Department,
Fermi National Accelerator Laboratory \\[-0.3cm]
P.O.Box 500, Batavia, IL 60510, USA\\[-0.3cm]
parke@fnal.gov
}

\maketitle

\vspace{-0.7cm}
\hfuzz=25pt
\tightenlines
\begin{abstract}
We present a general formalism for extracting information on the
fundamental parameters associated with neutrino masses and mixings
from two or more long baseline neutrino oscillation experiments.
This formalism is then applied to the current most likely experiments using
neutrino beams from the Japan Hadron Facility (JHF) and Fermilab's NuMI
beamline. 
Different combinations of muon neutrino or muon anti-neutrino
running are considered. 
To extract the type of neutrino mass hierarchy
we make use of the matter effect.
Contrary to naive expectation, we find that
both beams using neutrinos is more
suitable for determining the hierarchy
provided that the neutrino energy divided by baseline ($E/L$)
for NuMI is smaller than or equal to that of JHF.
Whereas to determine the small mixing angle, $\theta_{13}$, 
and the CP or T violating phase $\delta$, one 
neutrino and the other anti-neutrino is most suitable.
We make extensive use of bi-probability diagrams for both understanding
and extracting the physics involved in such comparisons. 
\end{abstract}
\newpage

\section{Introduction}
The solar neutrino puzzle, which has lasted nearly 40 years, 
has finally been resolved by KamLAND \cite{KamLAND} 
after invaluable contributions by numerous solar neutrino experiments,
especially recently by SNO\cite{SNO} and Super-Kamiokande\cite{SK}. 
The Large-Mixing-Angle (LMA) region of the
Mikheev-Smirnov-Wolfenstein (MSW) \cite{MSW} 
triangle \cite{triangle} has now been uniquely selected.
This resolution to the solar neutrino puzzle
has opened the door to the experimental exploration of 
CP violation in the Lepton Sector.
Together with the existing evidence for neutrino oscillation 
that has been obtained by the pioneering atmospheric neutrino 
observations \cite{SKatm} and the long-baseline accelerator 
experiments \cite{K2K}, it lends further support to the 
standard three-flavor mixing scheme of neutrinos. 

The remaining task toward the goal of uncovering the complete 
structure of the lepton flavor mixing would be a determination 
of the (1-3) sector of lepton mixing matrix
which is now called as the Maki-Nakagawa-Sakata 
matrix (MNS)\cite{MNS}.
A full determination of the (1-3) sector 
includes a determination of $\theta_{13}$, the CP or T 
violating phase $\delta$
and the sign of $\Delta m^2_{13}$. 
(We define 
$\Delta m^2_{ij} \equiv m^2_j - m^2_i$, where 
$m_i$ is the mass of the $i$-th eigenstate.)

The sign of $\Delta m^2_{13}$ signals which pattern of neutrino
masses nature has chosen, 
the normal hierarchy, $\Delta m^2_{13} > 0$, or the 
inverted hierarchy, $\Delta m^2_{13} < 0$.
This hierarchy must certainly carry interesting information 
on the, yet undiscovered, underlying principle of 
how nature organizes the neutrino sector. 
While the hierarchy question is of great interest, at this moment,
there is no experimental information 
available.
One possible exception is a hint from the neutrino data from 
supernova SN1987A \cite{MN01sn}, 
however the basis of this hint is under discussion\cite{raffelt02}.

To design the experiments to measure CP or T violating phase $\delta$ 
we must know in advance $\theta_{13}$, or at least its order of 
magnitude. There have been many proposals for experiments which 
may be able to measure $\theta_{13}$. They can be classified 
into two categories, long-baseline (LBL) accelerator experiments 
\cite {MINOS,OPERA,JHF,NuMI,SPL}, 
and reactor experiments \cite {krasnoyarsk,kashkari}.
While $\theta_{13}$ may be eventually measured by some of 
these experiments it was recognized that this measurement 
suffers from a problem of parameter degeneracy 
\cite{degeneracy,Burguet-C,MNjhep01,KMN02,BMW01,MNP2}. 
It stems from the fact that the determination of $\theta_{13}$, 
$\delta$ and the sign of $\Delta m^2_{13}$ are inherently 
coupled with each other.

A complete solution to the parameter degeneracy would require 
extensive measurement at two different energies and/or two 
different baselines
\cite{Burguet-C,BMW02a,BurguetC2},
or to combine two different experiments
\cite{kashkari,BurguetC2,DMM02,BMW02b,HLW02}.
We, however, seek an alternative strategy in this paper. 
That is, we propose to solve it one by one. 
We pursue the possibility that determination of the sign of 
$\Delta m^2_{13}$ may be carried out with only a limited set 
of measurement.
It is similar in spirit to the one employed in \cite{KMN02} 
in which we proposed a way to circumvent at least partly 
the problem of ($\theta_{13}$, $\delta$) degeneracy.

In this paper, we explore the line of thought of comparing
measurement at two different energies and/or combining two
different experiments to determine the sign of 
$\Delta m^2_{13}$ and $\theta_{13}$. 
In particular, it is of concern to us
how to combine the JHF \cite{JHF} and NuMI Off-Axis \cite{NuMI}
experiments. Apart from possibility of earlier detection
in MINOS, OPERA or reactor experiments, the
effect of nonzero $\theta_{13}$ is likely to be determined by
observing electron appearance events in these LBL experiments.
Hence, it is of great importance to uncover ways by which 
these two experiments can compliment one another.
We discuss in this paper which combination of modes of operation
($\nu$ or $\bar{\nu}$ channels) and which neutrino energies 
divided by baseline ($E/L$) 
have better sensitivity for determining the sign of $\Delta m^2_{13}$ and/or
measuring $\theta_{13}$.

While our treatment applies to wider context beyond application to 
the particular experiments, finding a way of optimizing the JHF-NuMI 
interplay is, we believe, a very relevant and urgent question for both 
of these projects.  
It is likely that the JHF experiment in its phase I operates in 
neutrino mode \cite{JHF}. 
Given the state of the other projects, NuMI is the one which is most 
likely to become a reality.
Therefore, it is important to find the most 
profitable mode of operation of the NuMI Off-Axis project.

We start by giving a general formalism of comparing two 
different measurement. 
It is to illuminate the analytic structure of the problem 
of two experiments comparison, and it would be of help in 
understanding the physical characteristics behind the observations 
we will provide. 
Instead of engaging an intensive $\chi^2$ analysis we 
prefer to illuminate the global chart of the two experiment 
comparison. 
To carry this out we rely heavily on a generalized 
version of the CP/T bi-probability plot that was introduced 
and developed in Refs.~\cite{MNjhep01,MNP1,MNP2}.
Our discussion in this paper,
 while not the complete story,
it is sufficient to enable us to understand the key physics points.
The experimental issues that need to be addressed to complete
the picture are beyond the scope of this paper.

The reader who wants to focus on the physics conclusions, 
in particular on our key observations in JHF-NuMI comparison, 
can go directly to Sec.~III, skipping Sec.~II. 
Then, they might want to come back to read the Sec.~II 
for a deeper understanding.

\section{General formalism for comparison of two different measurement}

\def \J{\alpha}
\def \N{\beta}

Let us combine two measurements, either $\nu$ or $\bar{\nu}$, 
for two different energies and/or baselines. For example, 
one measurement could come from JHF and 
another measurement from NuMI, 
with both $\nu$ (or $\bar{\nu}$) or one $\nu$ and the other 
$\bar{\nu}$.
All combinations can be treated in the universal formalism presented here.

We can start with two basic equations 
as in \cite{MNP2} for small $\sin \theta_{13}$: 
\begin{eqnarray}
P^{\J}(\nu) &=& X^\J \theta^2 + 
Y^\J \theta \cos {\left( \delta + \frac{\Delta^\J}{2} \right)} + 
P^\J_{\odot} \nonumber \\
P^{\N}(\nu) &=& X^\N \theta^2 + 
Y^\N \theta 
\cos {\left( \delta + \frac{\Delta^\N}{2} \right)} + P^\N_{\odot} 
\label{basicequations}
\end{eqnarray}
where $\theta \equiv \sin \theta_{13}$, and 
the superscripts $\J$ and $\N$ label the process and the experimental
setup. 
The value of the parameters 
$X^\J$, $Y^\J$ and $\Delta^\J$ depend on the process (specified by 
$\alpha$), the energy and path length 
of the experiment as well as sign of $\Delta m^2_{13}$ and the density
of matter traversed by the neutrino beam.
In Table I we give the values for the $X$'s, $Y$'s and $\Delta$'s
in terms of the variables $X_{\pm}$, $Y_{\pm}$ and  $\Delta_{13}$ 
which will be defined below. 

\begin{center}
\begin{table}
\begin{tabular}{|c||c|c||c|c||c|c||c|c||}
\hline
Process &$ \nu_\mu \rightarrow \nu_e $ &$ \nu_\mu \rightarrow \nu_e $ 
& $\bar{\nu}_\mu \rightarrow \bar{\nu}_e $ &
$ \bar{\nu}_\mu \rightarrow \bar{\nu}_e $ &
$ \nu_e \rightarrow \nu_\mu $ &$ \nu_e \rightarrow \nu_\mu $ 
& $\bar{\nu}_e \rightarrow \bar{\nu}_\mu $ &
$ \bar{\nu}_e \rightarrow \bar{\nu}_\mu $ 
\\ \hline \hline
 sign $\Delta m^2_{13}$ & $+$ ve & $-$ ve &$+$ ve &$-$ ve &$+$ ve &$-$ ve 
&$+$ ve &$-$ ve \\
\hline
$X$ & $X_+$ &$X_-$ &$X_-$ &$X_+$ &$X_+$ &$X_-$ &$X_-$ &$X_+$ \\ \hline
$Y$& $Y_+$ & $Y_-$ & $-Y_-$ & $-Y_+$ & $Y_+$ & $Y_-$ & $-Y_-$ & $-Y_+$\\ \hline
$\Delta$ &$\Delta_{13}$ &$-\Delta_{13}$ &$-\Delta_{13}$ &$\Delta_{13}$ &
$-\Delta_{13}$ &$\Delta_{13}$ &$\Delta_{13}$ &$-\Delta_{13}$ \\
\hline
\end{tabular}
 
\caption[]{The assignments for the parameters of 
Eq.(\ref{basicequations})
depending on the oscillation processes and sign of $\Delta m^2_{13}$.}
\end{table}
\end{center}
The functions $X_{\pm}$, $Y_{\pm}$ and $P_\odot$ are given by \cite{golden}
\begin{eqnarray}
X_{\pm} &=& 4 s^2_{23} 
\left(
\frac{\Delta_{13}}{B_{\mp}} 
\right)^2
\sin^2{\left(\frac{B_{\mp}}{2}\right)}, 
\label{X} \\
Y_{\pm} &=& \pm 8 c_{12}s_{12}c_{23}s_{23}
\left(
\frac{\Delta_{12}}{aL}
\right)
\left(
\frac{\Delta_{13}}{B_{\mp}}
\right)
\sin{\left(\frac{aL}{2}\right)}
\sin{\left(\frac{B_{\mp}}{2}\right)}
\label{Y}\\
P_{\odot} & = & c^2_{23} \sin^2{2\theta_{12}} 
\left(\frac{\Delta_{12}}{aL}\right)^2
\sin^2{\left(\frac{aL}{2}\right)}
\end{eqnarray}
with 
\begin{eqnarray}
\Delta_{ij}  \equiv \frac{|\Delta m^2_{ij}| L}{2E}
\quad
{\rm and} \quad B_{\pm} \equiv |\Delta_{13} \pm aL|,
\end{eqnarray}
where $a = \sqrt{2} G_F N_e$ denotes the index of refraction 
in matter with $G_F$ being the Fermi constant and $N_e$ a constant 
electron number density in the earth. 
Some crucial properties of the functions $X_{\pm}$ and $Y_{\pm}$ 
which allow this unified notation are summarized in 
Appendix A.

The basic equations (\ref{basicequations}) can be solved for 
either sign of $\Delta m^2_{13}$ as: 
\begin{eqnarray}
\theta & = & \sqrt{\frac{P^\J-P^\J_{\odot}}{X^\J}} - 
\frac{Y^\J}{2X^\J} 
\cos {\left(\delta + \frac{\Delta^\J}{2} \right)}
~=~
\sqrt{\frac{P^\N-P^\N_{\odot}}{X^\N}} - 
\frac{Y^\N}{2X^\N} 
\cos {\left( \delta + \frac{\Delta^\N}{2} \right)}.
\label{basicsolution}
\end{eqnarray}
where we have ignored terms of order $\frac{Y^2}{X}$.
The right equality can be used to solve for both $\sin \delta$ and 
$\cos \delta$ from which $\theta$ is then determined.
This is a straightforwardly generalization of the method used 
in \cite{MNP2} 
to obtain, analytically, the degenerate set of solutions
and to elucidate the relationships between them. 
Here, we will present the result of such a general analysis in 
the Appendix A. An important point, however, is that once the 
solutions for $\delta$ are known, $\theta$ is easily obtained 
using left hand equality in Eq.~(\ref{basicsolution}).

Experimentally the most likely channels to be realized in the near future
are $\nu_\mu \rightarrow \nu_e$ and $\bar{\nu}_\mu \rightarrow \bar{\nu}_e$,
therefore in the next two subsections we will specialize to these
channels with both beams being Neutrinos (or Anti-Neutrinos)
and one beam of Neutrinos and one beam of Anti-Neutrinos.
In the following section we will consider the specific experimental situations
presented by the JHF\cite{JHF} and NuMI\cite{NuMI} proposals.

\subsection{Both Neutrinos or Both Anti-Neutrinos}

We apply the general formalism developed in the previous section 
to an explicit example of a $\nu$-$\nu$ comparison
of two experiments with different energies and baseline distances. 
At the end of this subsection we
will give the relationship to the  $\bar{\nu}$-$\bar{\nu}$ comparison. 
We will label the two experiments J and N in reference to JHF and NuMI.
However, in spite of the reference to specific two projects, our general 
discussions are valid for any pair of the LBL experiments on the globe
and can also be used to discuss other processes but here we will
concentrate on  $\nu_\mu \rightarrow \nu_e$.

We start by introducing some convenient notation. 
Since we will treat $\Delta m^2_{13} > 0$ and\\ $\Delta m^2_{13} < 0$
simultaneously we will label the various quantities below 
with subscripts $\pm$ to indicate the sign of $\Delta m^2_{13}$. 
We define
\begin{eqnarray}
S{\pm}
&\equiv& 
\pm
\frac{Y^J_{\pm}}{2X^J_{\pm}} 
\sin {\left(\frac{\Delta_{13}^J}{2}\right)} 
\mp 
\frac{Y^N_{\pm}}{2X^N_{\pm}} 
\sin {\left(\frac{\Delta_{13}^N}{2}\right)}, \\ 
C_{\pm} 
&\equiv& 
- \frac{Y^J_{\pm}}{2X^J_{\pm}} 
\cos {\left(\frac{\Delta_{13}^J}{2}\right)} 
+ \frac{Y^N_{\pm}}{2X^N_{\pm}} 
\cos {\left(\frac{\Delta_{13}^N}{2}\right)}, \\
D_{\pm} 
&\equiv& 
S_{\pm}^2+C_{\pm}^2, \\
\Delta P_{\pm} 
&\equiv& 
\sqrt{\frac{P^J - P^J_{\odot}}{X^J_{\pm}}} - 
\sqrt{\frac{P^N - P^N_{\odot}}{X^N_{\pm}}}.
\label{JNnunu}
\end{eqnarray}
The signs and the subscript assignment come from Table 1.

By solving (\ref{basicsolution}) with these definitions, 
it is easy to derive (see Appendix A), for positive $\Delta m^2_{13}$, 
the allowed values of $(\theta,~\delta)$
which we assign subscripts 1 and 2;
\begin{eqnarray}
\sin{\delta_{1,2}} &=& 
\frac{1}{D_{+}}
\left[
-S_{+} \Delta P_{+}  \pm 
C_{+} \sqrt{D_{+} - (\Delta P_{+})^2}
\right], 
\nonumber \\
\cos{\delta_{1,2}} &=& 
\frac{1}{D_{+}}
\left[
-C_{+} \Delta P_{+} \mp 
S_{+} \sqrt{D_{+} - (\Delta P_{+})^2}
\right].
\label{delta+nunu}
\end{eqnarray}
Notice that with this choice of signs 
$\sin^2 \delta_i + \cos^2 \delta_i =1$.
For negative $\Delta m^2_{13}$, we use subscripts 3 and 4
to distinguish them from the positive $\Delta m^2_{13}$ solutions, 
\begin{eqnarray}
\sin{\delta_{3,4}} &=& 
\frac{1}{D_{-}}
\left[
-S_{-} \Delta P_{-}  \mp 
C_{-} \sqrt{D_{-} - (\Delta P_{-})^2}
\right],
\nonumber \\
\cos{\delta_{3,4}} &=& 
\frac{1}{D_{-}}
\left[
-C_{-} \Delta P_{-} \pm 
S_{-} \sqrt{D_{-} - (\Delta P_{-})^2}
\right].
\label{delta-nunu}
\end{eqnarray}
The relative $\pm$ signs in (\ref{delta+nunu}) and (\ref{delta-nunu}) 
are arbitrary at this point but are the same as those used in the next 
subsection.

The relationships between the mixed-sign degenerate solution 
are given by 
\begin{eqnarray}
\cos{(\delta_{1} - \delta_{3})} &=& 
\frac{\left(C_{+}C_{-} + S_{+}S_{-} \right)}{D_{+} D_{-}}
\left(\Delta P_{+} \Delta P_{-} - 
\sqrt{D_{+} - \Delta P_{+}^2} 
\sqrt{D_{-} - \Delta P_{-}^2}
\right)
\nonumber \\
&+& 
\frac{\left(C_{-}S_{+} - C_{+}S_{-} \right)}{D_{+} D_{-}}
\left(\Delta P_{-} 
\sqrt{D_{+} - \Delta P_{+}^2} 
+
\Delta P_{+}
\sqrt{D_{-} - \Delta P_{-}^2}
\right), 
\label{del-diffc} \\
\sin{(\delta_{1} - \delta_{3})} &=& 
\frac{\left(C_{-}S_{+} - C_{+}S_{-} \right)}{D_{+} D_{-}}
\left(\Delta P_{+} \Delta P_{-} - 
\sqrt{D_{+} - \Delta P_{+}^2} 
\sqrt{D_{-} - \Delta P_{-}^2}
\right)
\nonumber \\
&-& 
\frac{\left(C_{+}C_{-} + S_{+}S_{-} \right)}{D_{+} D_{-}}
\left(\Delta P_{-} 
\sqrt{D_{+} - \Delta P_{+}^2} +
\Delta P_{+}
\sqrt{D_{-} - \Delta P_{-}^2}
\right).
\label{del-diffs}
\end{eqnarray}

We note that, unlike the case of $\nu$-$\bar{\nu}$ comparison 
\cite {MNP1} which will be discussed in the next section, 
there is no sensible limit 
$\Delta^N_{13} \rightarrow \Delta^J_{13}$ and 
$(aL)^N \rightarrow (aL)^J$ because in this limit
no extra information is provided.
However, one can ask the question, can we obtain some useful  
relationships between CP phases of the mixed-sign degeneracy 
in an approximation of small difference in the matter effect 
between JHF and NuMI? The answer is yes and we will derive 
such relations in Appendix B where we formulate such 
perturbative treatment.

The solutions for $\theta$ can easily be obtained by substituting 
appropriate values of $\sin \delta$ and $\cos \delta$ into
Eq.(\ref{basicsolution}). 
For positive $\Delta m^2_{13}$, the values of $\theta_{1,2}$ are 
obtained by using the $X^J_+,~Y^J_+,~\Delta^J_{13}$ and
$X^N_+,~Y^N_+,~\Delta^N_{13}$ in Eq.(\ref{basicsolution}) and for 
negative  $\Delta m^2_{13}$, the values of $\theta_{3,4}$ are
obtained by using   
$X^J_-,~Y^J_-,~-\Delta^J_{13}$ and $X^N_-,~Y^N_-,~-\Delta^N_{13}$.
See Eq.(\ref{thetas}) of the Appendix A for explicit expressions.

For the $\bar{\nu}$-$\bar{\nu}$ comparison:
for positive $\Delta m^2_{13}$, the values of $\theta_{1,2}$ are 
obtained by using the $X^J_-,~-Y^J_-,~-\Delta^J_{13}$ and
$X^N_-,~-Y^N_-,~-\Delta^N_{13}$ in Eq.(\ref{basicsolution}) and for 
negative  $\Delta m^2_{13}$, the values of $\theta_{3,4}$ are
obtained by using   
$X^J_+,~-Y^J_+,~\Delta^J_{13}$ and $X^N_+,~-Y^N_+,~\Delta^N_{13}$.
Thus the allowed region for the $\bar{\nu}$-$\bar{\nu}$ comparison  is  
identical to the 
$\nu$-$\nu$ comparison except for the fact that the 
roles of $\Delta m^2_{13} >0 $ and
$\Delta m^2_{13} <0$ are interchanged.\footnote{The difference in the 
sign of the Y coefficients can be compensated by taking 
$\delta \rightarrow \delta +\pi$, see Eq.(\ref{basicequations}).
Thus the allowed regions are identical.}

\subsection{One Neutrinos, One Anti-Neutrinos}

We want to treat the $\nu$-$\bar{\nu}$ comparison in an entirely 
analogous fashion as the $\nu$-$\nu$ comparison in the previous 
subsection. Toward this goal we introduce a similar notation:
\begin{eqnarray}
S^\prime_{\pm}
&\equiv& 
\pm
\frac{Y^J_{\pm}}{2X^J_{\pm}} 
\sin {\left(\frac{\Delta_{13}^J}{2}\right)} 
\mp 
\frac{Y^N_{\mp}}{2X^N_{\mp}} 
\sin {\left(\frac{\Delta_{13}^N}{2}\right)}, \\ 
C^\prime_{\pm} 
&\equiv& 
- \frac{Y^J_{\pm}}{2X^J_{\pm}} 
\cos {\left(\frac{\Delta_{13}^J}{2}\right)} 
- \frac{Y^N_{\mp}}{2X^N_{\mp}} 
\cos {\left(\frac{\Delta_{13}^N}{2}\right)}, \\
D^\prime_{\pm} 
&\equiv& 
(S^\prime_{\pm})^2 + (C^\prime_{\pm})^2, \\
\Delta P^\prime_{\pm} 
&\equiv& 
\sqrt{\frac{P^J - P^J_{\odot}}{X^J_{\pm}}} - 
\sqrt{\frac{P^N - P^N_{\odot}}{X^N_{\mp}}}.
\label{JNmixed}
\end{eqnarray}
where the subscript $\pm$ in (\ref{JNmixed}) denote the sign 
of $\Delta m^2_{13}$ and here $P^N$ is for anti-neutrinos. 
The signs and the subscript assignment come from Table 1.
These definitions are almost identical to the $\nu$-$\nu$ definitions
but not quite, the $\pm$ labels on the left hand side are different
and these difference lead to markedly different outcomes. 

It is easy to verify that by using the above notation 
$\sin{\delta_{i}}$ as well as $\theta_{i}$ are given by 
exactly the same expressions as in the previous sections, e.g.,  
(\ref{delta+nunu}) and (\ref{delta-nunu}), 
apart from replacement 
$S_{\pm} \rightarrow S^\prime_{\pm}$, 
$C_{\pm} \rightarrow C^\prime_{\pm}$, and so on. 

The relative $\pm$ signs in (\ref{delta+nunu}) and (\ref{delta-nunu}) 
are determined such that it reproduces the pair of the 
degenerate solutions, $\delta_{3} = \delta_{1} + \pi$ (mod. $2\pi$) 
in the limit of 
$P^N=P^J$, $\Delta^{J} = \Delta^{N}$ and $(aL)^J = (aL)^N$  
for a given value of $\theta_{13}$ \cite {MNP2}.
The relation, however, receives corrections due to difference 
in matter effect between JHF and NuMI even at $\Delta^J=\Delta^N$. 
It is not difficult to compute the correction 
at $P^N = P^J$ in the form 
\begin{eqnarray}
\delta_{3} = \delta_{1} + \pi + 
C^{N-J} \left[(aL)^N - (aL)^J\right],
\label{mattcorr}
\end{eqnarray}
to first order in JHF-NuMI difference of matter effects.
The coefficient $C^{N-J}$ is given in Appendix B.

Notice that the relation between the same-sign solutions of 
the CP violating phase is given generically by the formula 
(\ref{deltadiff}) in comparison of any two channels.
Among other things, $\delta_{2} = \pi - \delta_{1}$ always 
holds at oscillation maximum, $\Delta^N = \Delta^J = \pi$, 
despite the difference in matter effects at JHF and NuMI.
It must be the case since the bi-probability trajectories shrink 
to straight lines.\footnote
{The straight-line CP trajectory can be achieved even when 
energy distribution of neutrino flux times cross section is 
taken into account \cite{KMN02}.
}

\section{JHF verses NuMI Comparison}

In this section we will discuss explicitly the JHF and NuMI 
possibilities. JHF will have a baseline of 295km, and the   
neutrino energy at which oscillation maximum occurs is
\begin{equation}
0.60 ~\text{GeV} \left({\Delta m^2_{13} \over 2.5 \times 10^{-3} ~eV^2}\right).
\end{equation}
For our figures we will use this energy plus 0.8 GeV which is 
33\% higher and near the
peak in the event rate for $\nu_e$ appearance due to the rising cross 
section but falling oscillation probability assuming the same flux can 
be maintained.

So far the path length for NuMI is undecided but it will most likely
be between 500 and 1000 km. 
We use 732km for our figures, the path length of NuMI/MINOS.
At this distance oscillation maximum occurs at 
\begin{equation}
1.5 ~\text{GeV} \left({\Delta m^2_{13} \over 2.5 \times 10^{-3} ~\text{eV}^2}\right)
\left({L \over 732 ~\text{km}}\right).
\end{equation}
The other energy used is 33\% higher at 2.0 GeV near the event rate peak
for the reasons stated for JHF. 

In Fig.~1 we have plotted the allowed region in bi-probability space 
assume no knowledge of $\theta$ and $\delta$
in a comparison between JHF neutrinos and NuMI neutrinos. This allowed
region forms two narrow ``pencils'' which originate at the origin 
and grow in width away from the origin.
Associated with each of these ``pencils'' is the sign of $\Delta m^2_{13}$,
thus if the pencils are well separated then a comparison of these
two experiments can be used to determine the sign of $\Delta m^2_{13}$.
But the size of $\theta_{13}$ is poorly determined in such a comparison.
More details on this $\nu$-$\nu$ comparison will be given in the next
subsection. (A $\bar{\nu}$-$\bar{\nu}$ comparison is identical but with 
the sign of $\Delta m^2_{13}$ flipped.)

In Fig.~2 we have plotted the allowed region in bi-probability
space in a comparison between JHF neutrinos and NuMI anti-neutrinos.
Here the allowed regions are very broad and there is significant overlap
between the allowed regions for the two signs of $\Delta m^2_{13}$.
However, the size of $\theta_{13}$ can be determined with reasonable accuracy
in such a comparison. 
Fig.~3 is similar to Fig.~2 but here the concentration is on the
overlap region. 
More details of this $\nu$-$\bar{\nu}$ will be given in the 
subsection following the $\nu$-$\nu$ comparison.
(A $\bar{\nu}$-$\nu$ comparison is identical to a  $\nu$-$\bar{\nu}$ but with 
the sign of $\Delta m^2_{13}$ flipped.)

\subsection{Both Neutrinos}

It is worthwhile to have a simple formula which relates 
between two $\nu_{\mu} \rightarrow \nu_{e}$ appearance 
probabilities obtained by JHF and NuMI both in neutrino 
(or anti-neutrino) channel. Let us restrict, for definiteness, 
the following discussion into the case of neutrino channel.  
Then, one can easily derive the following expression 
\begin{eqnarray}
\frac{P^N - P^N_{\odot}}{X^N_{\pm}} - 
\frac{P^J - P^J_{\odot}}{X^J_{\pm}} &=&  
~~\left[
\frac{Y^N_{\pm}}{X^N_{\pm}} 
\cos {\left(\frac{\Delta_{13}^N}{2}\right)} 
- \frac{Y^J_{\pm}}{X^J_{\pm}} 
\cos {\left(\frac{\Delta_{13}^J}{2}\right)} 
\right] 
\theta \cos{\delta} 
\nonumber \\
& & -
\left[
\frac{Y^N_{\pm}}{X^N_{\pm}} 
\sin {\left(\frac{\Delta_{13}^N}{2}\right)} - 
\frac{Y^J_{\pm}}{X^J_{\pm}} 
\sin {\left(\frac{\Delta_{13}^J}{2}\right)} 
\right] 
\theta \sin{\delta} 
\nonumber \\
&=&
2 
\left(
C_{\pm} \cos{\delta} \pm S_{\pm} \sin{\delta} 
\right)
\theta
\label{P^J-P^N}
\end{eqnarray}

We want to understand the behavior of two loci corresponding to 
positive and negative $\Delta m^2_{13}$ in Fig.~1 as a function of 
$\Delta^J$ and $\Delta^N$.
In particular, we are interested in how the slopes of the two
``pencils'' change in Fig.~1. To this goal we compute the slope ratio 
of the central axis of positive $\Delta m^2_{13}$ to the 
negative $\Delta m^2_{13}$. 
Since the central axis of the ``pencil'' is obtained by equating 
LHS of Eq.~(\ref{P^J-P^N}) to zero, the slope of $P^N-P^J$ line 
in Fig.~1 is simply given by 
$\alpha_{\pm} \equiv \frac{X^N_{\pm}}{X^J_{\pm}}$. 
Thus the slope ratio can be written as
\begin{eqnarray}
\frac{\alpha_{+}}{\alpha_{-}} =  
\frac{\frac{X^N_{+}}{X^N_{-}}} {\frac{X^J_{+}}{X^J_{-}}}
 =
\frac{
\sin^2{\left(\frac{B_{-}^N}{2} \right)}/(\frac{B_{-}^N}{2})^2
}{
\sin^2{\left(\frac{B_{+}^N}{2} \right)}/(\frac{B_{+}^N}{2})^2
}
\times 
\frac{
\sin^2{\left(\frac{B_{+}^J}{2} \right)}/(\frac{B_{+}^J}{2})^2
}{
\sin^2{\left(\frac{B_{-}^J}{2} \right)}/(\frac{B_{-}^J}{2})^2
}
\label{slratio}
\end{eqnarray}
Therefore, the behavior of the slope ratio is controlled by 
a single function $\sin^2{x}/x^2$. 

To have a qualitative understanding we perform perturbation 
expansion assuming matter effect is small, $aL \ll \Delta_{13}$.
Noting that 
$B_{\pm} = \Delta_{13} \pm aL$ for both JHF and NuMI 
we obtain, to first order in matter effect,  
\begin{eqnarray}
\frac{\alpha_{+}}{\alpha_{-}} = 1 + 
2 \left[
\frac{2}{\Delta_{13}^N} - \cot{\left(\frac{\Delta_{13}^N}{2} \right)}
\right] (aL) |_{NuMI} - 
2 \left[
\frac{2}{\Delta_{13}^J} - \cot{\left(\frac{\Delta_{13}^J}{2} \right)}
\right] (aL) |_{JHF}
\label{app_slratio}
\end{eqnarray}
Therefore, the slope of positive-$\Delta m^2_{13}$ ``pencil'' 
is larger than negative-$\Delta m^2_{13}$ ``pencil'' because of the 
larger matter effect in NuMI.

To understand the energy dependence of the slope ratio we 
note that $\frac{1}{x} - \cot{x}$ is a monotonically increasing 
function of $x$ in $0 \leq x \leq \pi$.
Let us first examine the case of $\Delta_{13}^N = \Delta_{13}^J$. 
When the neutrino energy is increased from the oscillation maximum 
(which means lowering $\Delta_{13}$'s) the slope ratio decreases, 
in agreement with the behavior in Fig.~1, from 1(a) to 1(d). 
We have checked that lowering the energy (larger $\Delta_{13}$'s) 
in fact leads to larger slope ratio.  However at energies below oscillation
maximum the probabilities and event rates rapidly get smaller.

The behavior of slopes of ``pencils'' in other parts in Fig.~1 
can also be understood in a similar manner. From Fig.~1(a) to 1(b) 
the slope ratio decreases, whereas from Fig.~1(a) to 1(c) it increases. 
The change in 1(a) to 1(b) is obtained by lowering $\Delta_{13}^N$ 
while keeping $\Delta_{13}^J$ fixed. Since the second term in 
the RHS of (\ref{app_slratio}) decreases as $\Delta_{13}^N$ 
decreases, and hence the slope ratio decreases, in agreement 
with Figs.~1(a) and 1(b). Similarly, from Fig.~1(a) to 1(c) 
$\Delta_{13}^J$ decreases while keeping $\Delta_{13}^N$ is 
kept fixed. It should lead relatively small amount of increase 
in the slope ratio, again in agreement with Figs.~1(a) and 1(c). 

However the slope of these ``pencils'' is not the total story, the width
of the ``pencils'' is also important. Although the ratio of slopes is
slightly larger in Fig.~1(c) than Fig.~1(a) the width of the ``pencils'' is
significantly larger for Fig.~1(c) than Fig.~1(a)
such that the separation of the allowed regions is smaller for Fig.~1(c) than 
Fig.~1(a).  
The square of the width of the ``pencils'' is controlled by the quantity
\begin{eqnarray}
4(S_{\pm}^2 + C_{\pm}^2)  & = & 
\left( \frac{Y^N_{\pm}}{X^N_{\pm}} \right) ^2
+
\left( \frac{Y^J_{\pm}}{X^J_{\pm}} \right) ^2
-
2\left( \frac{Y^N_{\pm}}{X^N_{\pm}} \right)
\left( \frac{Y^J_{\pm}}{X^J_{\pm}} \right)
\cos ( \frac{\Delta^N_{13}}{2} - \frac{\Delta^J_{13}}{2} ).
\end{eqnarray}  
For $\Delta^N_{13} = \Delta^J_{13}$, the width equals
\begin{eqnarray}
\left( \frac{Y^N_{\pm}}{X^N_{\pm}} \right.
& - & \left. \frac{Y^J_{\pm}}{X^J_{\pm}} \right)^2,
\end{eqnarray}
and is very small. 
This smallness follows from the fact that, at the same $\Delta_{13}$,
the identity  
$\frac{Y^N_{\pm}}{\sqrt{X^N_{\pm}}}  = \frac{Y^J_{\pm}}{\sqrt{X^J_{\pm}}}$
holds ( $\frac{Y}{\sqrt{X}}$ depends only on vacuum parameters),
and ratio of $X^N_{\pm}$ to ${X^J_{\pm}}$ is close to unit.
If $\Delta^N_{13} \neq \Delta^J_{13}$, then 
$\frac{Y^N_{\pm}}{\sqrt{X^N_{\pm}}} \neq \frac{Y^J_{\pm}}{\sqrt{X^J_{\pm}}}$
and the width of the pencil grows rapidly as the cancellation 
that occurs for the same $\Delta_{13}$ no longer holds. 
Thus the width of the ``pencils'' in Fig.~1(b) and Fig.~1(c) 
are approximately equal and are much larger than the width 
in Fig.~1(a) and Fig.~1(d).

The conclusion to be gained from these results and figures is 
that if you are making a comparison of a
JHF neutrino experiment and a NuMI neutrino experiment to determine
the sign $\Delta m^2_{13}$ then the best separation occurs
when $\Delta^J = \Delta^N$ i.e. the same $\frac{E}{L}$ for 
both experiments. Smaller values of the $E$ are slightly preferred 
so that we expect the optimum value, once all experimental issues 
are included, to be near oscillation maximum (Fig.~1(a) and 1(d)).
Away from the same  $\frac{E}{L}$ for both experiments
the separation between the allowed region becomes worse,
apart from special ranges of the CP or T violating phase $\delta$ when
$\left(\frac{E}{L}\right)|_{JHF} > \left(\frac{E}{L}\right)|_{NuMI}$ 
(Fig.~1(c)).
Whereas for the choice 
$ \left(\frac{E}{L}\right)|_{JHF} < \left(\frac{E}{L}\right)|_{NuMI}$
(i.e $\Delta^J > \Delta^N $) there is significant overlap between the two
different sign of $\Delta m^2_{13}$ allowed regions (Fig.~1(b)).
(If one moves further, say, to $E_{JHF} = 0.6$ GeV and 
$E_{NuMI} = 2.5$ GeV, the two pencils overlap almost completely.)
In these regions no matter how accurate the experiment the sign of 
$\Delta m^2_{13}$ cannot be determined by this comparison.
(For JHF anti-neutrinos versus NuMI anti-neutrinos the conclusion is
the same.)

In summary, for the comparison JHF neutrinos to NuMI neutrinos to be useful
in the determination of the sign of $\Delta m^2_{13}$, choosing
\begin{eqnarray}
\left( \frac{E}{L}\right)|_{JHF} \geq \left( \frac{E}{L} \right) |_{NuMI}
\end{eqnarray}
is of great importance with the preference for 
equal $\left(\frac{E}{L}\right)$.
For $\left( \frac{E}{L}\right) |_{JHF} 
< \left( \frac{E}{L}\right)|_{NuMI}$ there is significant
overlap between the narrow allowed regions so that this is not
a good choice for this comparison.

Contrary to advantage of $\nu$-$\nu$ comparison in 
determination of the sign of $\Delta m^2_{13}$, measurement of 
$\theta_{13}$ suffers from large uncertainty in this channel.
The bi-probalitity trajectory, for a given $\theta$, moves nearly along the 
direction of the ``pencil'' as one varies $\delta$, thus
any measurement of the oscillation probabilities 
with finite resolution cannot pinpoint the value of $\theta$.
To indicate this point we have plotted in Fig.~1 by numbers in \% 
beside thin arcs the fractional difference
$\Delta \theta/\theta 
\equiv (\theta_2 - \theta_1)/\left((\theta_2 + \theta_1)/2 \right)$, 
where $\theta_2$ and $\theta_1$ here indicate the maximal and 
the minimal values of $\theta_{13}$ along each arc. The numbers 
are large, typically 30 \% or even larger. 

In fact, it is easy to do analytical estimate of 
the fractional difference. Suppose that we have measured an 
appearance probability $P^J$ or $P^N$. We assume for simplicity 
that we know the sign of $\Delta m^2_{13}$, and restrict 
ourselves into the case of oscillation maximum, 
$\Delta^J_{13}=\Delta^N_{13}=\pi$. 
The measured probability allows a range of $\theta_{13}$ 
\cite{KMN02} 
and one can easily calculate the maximal value 
of $\Delta \theta/\theta$ within the range, which leads to
\begin{eqnarray}
\frac{\Delta \theta}{\theta} = 
\sin{2 \theta_{12}} c_{23}
\frac{\Delta_{12}}{\sqrt{P^{J/N}}}
\left(
\frac{\sin{(aL/2)}}{aL/2}
\right).
\label{dtheta/theta}
\end{eqnarray}
Since $P^{J/N}$ is a few \% level, it leads to a few tens 
in \% of the fractional difference, in agreement with 
the numbers in Fig.~1.

\subsection{One Neutrinos and One Anti-Neutrinos}

We now turn to the comparison for JHF neutrinos and NuMI anti-neutrinos
or vice versa with flipped signs of $\Delta m^2_{13}$.
Again one can write a simple formulae relating the two 
oscillation probabilities similar to Eq.~(\ref{P^J-P^N}) 
\begin{eqnarray}
\frac{P^N - P^N_{\odot}}{X^N_{\mp}} - 
\frac{P^J - P^J_{\odot}}{X^J_{\pm}} &=&  
~~\left[
\frac{- Y^N_{\mp}}{X^N_{\mp}} 
\cos {\left(\frac{\Delta_{13}^N}{2}\right)} 
- \frac{Y^J_{\pm}}{X^J_{\pm}} 
\cos {\left(\frac{\Delta_{13}^J}{2}\right)} 
\right] 
\theta \cos{\delta} 
\nonumber \\
& & -
\left[
\frac{ Y^N_{\mp}}{X^N_{\mp}} 
\sin {\left(\frac{\Delta_{13}^N}{2}\right)} - 
\frac{Y^J_{\pm}}{X^J_{\pm}} 
\sin {\left(\frac{\Delta_{13}^J}{2}\right)} 
\right] 
\theta \sin{\delta} 
\nonumber \\
&=&
2 \left(
C_{\pm} \cos{\delta} \pm S_{\pm} \sin{\delta} 
\right)\theta
\label{P^J-Pbar^N}
\end{eqnarray}
The square of the RHS of this equation controls the square of the width of the
allowed region and is given by
\begin{eqnarray}
4(S_{\pm}^2 + C_{\pm}^2)  & = & 
\left( \frac{Y^N_{\mp}}{X^N_{\mp}} \right) ^2
+
\left( \frac{Y^J_{\pm}}{X^J_{\pm}} \right) ^2
+
2\left( \frac{Y^N_{\mp}}{X^N_{\mp}} \right)
\left( \frac{Y^J_{\pm}}{X^J_{\pm}} \right)
\cos ( \frac{\Delta^N_{13}}{2} + \frac{\Delta^J_{13}}{2} )
\end{eqnarray}  
apart from an overall factor of $\theta^2$.
As $(\Delta^N_{13} + \Delta^J_{13})$ varies from 
$\pi$ (both experiments at energies twice the oscillation maximum energy)
to $2\pi$ (both experiments at the oscillation maximum energy) 
this squared width grows from
\begin{eqnarray}
\left( \frac{Y^N_{\mp}}{X^N_{\mp}} \right)^2
 +  \left( \frac{Y^J_{\pm}}{X^J_{\pm}} \right)^2
\quad & {\rm to} & \quad
\left( \frac{Y^N_{\mp}}{X^N_{\mp}} \right.
 -  \left. \frac{Y^J_{\pm}}{X^J_{\pm}} \right)^2.
\end{eqnarray}
Note that the later width is larger than the first width 
since $Y_-/X_-$ has opposite sign to $Y_+/X_+$
(no cancellation occurs here).
Thus the width of the allowed region grows as you lower the energies of the
experiments to oscillation maximum.  This can be seen in Fig.~2; 2(d) has the
smallest width whereas 2(a) has the largest in accordance with the above 
statement.

There is significant overlap between the allowed regions 
of positive and negative $\Delta m^2_{13}$ contours. 
Hence, the $\nu$-$\bar{\nu}$ comparison does not appear 
to be the right way for determining the sign, unless 
$\delta$ turns out to be close to $3 \pi/2$ and $\pi/2$ 
for $\Delta m^2_{13} > 0$ and $\Delta m^2_{13} < 0$ cases, 
respectively \cite{MNjhep01}.
However, this comparison may be better to measure $\theta_{13}$ 
if JHF operate only in neutrino channel, as planned, 
and if NuMI Off-Axis is running with significant overlap with JHF phase I.
For this purpose, the best way of measuring $\theta_{13}$ would be 
to tune the energy at oscillation maxima for both JHF and NuMI.
This is just an extension of the KMN strategy\cite{KMN02}.

The reader may be curious as to why the slopes of the 
shrunken trajectories at $\Delta_{13} = \pi$ are positive in 
$\nu$-$\nu$ comparison whereas they are negative in 
$\nu$-$\bar{\nu}$ comparison. This can be easily understood 
as follows:
The slopes of shrunken trajectories in $\nu$-$\nu$ comparison 
are given by 
$Y^N_{+}/Y^J_{+}$ ($Y^N_{-}/Y^J_{-}$) for positive (negative) 
$\Delta m^2_{13}$. 
On the other hand, the slopes in $\nu$-$\bar{\nu}$ comparison 
are given by 
$Y^N_{-}/Y^J_{+}$ ($Y^N_{+}/Y^J_{-}$) for positive (negative) 
$\Delta m^2_{13}$. 
Since $Y_{\pm}$ differ in sign, as in Eq.~(\ref{Y}), 
the latter slopes are negative definite, whereas the former 
are positive definite.
The fact that the $\nu$-$\nu$ trajectory is along the ``pencil'' 
and $\nu$-$\bar{\nu}$ trajectory is perpendicular to axis of the
``cigar'' can be understood by a similar argument.

\section{Another Interesting Comparison}

When we have access to $\nu_e$ and $\bar{\nu}_e$ beams then a
comparison of a $\nu_\mu \rightarrow \nu_e$ experiment with 
$\bar{\nu}_e \rightarrow \bar{\nu}_\mu$ experiment will be 
possible.\footnote{Possible ways to produce electron neutrino beams 
include muon storage rings (Neutrino Factories \cite{nufac}) 
and $\beta$-beams \cite{betabeams}.} 
This comparison is interesting because it directly compares two CPT
conjugate processes. 
If CPT is conserved, then at the same E/L the only difference between
the oscillation probabilities for $\nu_\mu \rightarrow \nu_e$ and
$\bar{\nu}_e \rightarrow \bar{\nu}_\mu$ can come from matter effects.
Thus this comparison is even more sensitive to the mass
hierarchy, i.e. the sign of $\Delta m^2_{13}$, than the neutrino-neutrino
comparison discussed earlier in the previous section.

For this comparison the bi-probability figure looks similar to Fig.~1
but the difference in slope between the ``pencils'' is enhanced
because the sign of the matter effect is opposite for
$\bar{\nu}_e \rightarrow \bar{\nu}_\mu$ compared to 
$\nu_\mu \rightarrow \nu_e$ and the CP or T violating terms are identical
in sign and magnitude.
Therefore this comparison provides an excellent way to 
separate the two signs of $\Delta m^2_{13}$.
If we label the two experiments as J and N as before then the 
ratio of slopes is given by
\begin{eqnarray}
\frac{\alpha_{+}}{\alpha_{-}} =  
\frac{\frac{X^N_{+}}{X^N_{-}}} {\frac{X^J_{-}}{X^J_{+}}}
 & = & 1 + 
2 \left[
\frac{2}{\Delta_{13}^N} - \cot{\left(\frac{\Delta_{13}^N}{2} \right)}
\right] (aL) |_{N} + 
2 \left[
\frac{2}{\Delta_{13}^J} - \cot{\left(\frac{\Delta_{13}^J}{2} \right)}
\right] (aL) |_{J}.
\label{app_slratio2}
\end{eqnarray}
This expression differs from that of Eq.(\ref{app_slratio}) by the sign of
the third term. 
Here the matter effects for both experiments enhance the ratio of the slopes
whereas in the neutrino-neutrino comparison early there was a partial
cancellation.
Although at present the comparison of this section is purely academic
it is instructive and useful for understanding the general nature of
the comparison of neutrino oscillation experiments and will be important
to test the possibility of CPT violation in the future.

\section{Summary and Conclusions}
We have presented a general formalism for comparing two or more
long baseline neutrino oscillation experiments and applied this 
formalism to the JHF and NuMI experiments.
The combination of modes that will be important depends on the question
one is asking and what other information is available at the time.
The use of bi-probability diagrams like the ones presented in this paper will
be important for understanding the physics issues and making trade offs 
at the time decisions are made regarding neutrino energies
and modes of operation.

In general, the both neutrino or both anti-neutrino comparison
is useful for determining whether the mass hierarchy is normal
or inverted, i.e. the sign $\Delta m^2_{13}$. 
However, it is important here that the experiment with the larger matter
effect (longer baseline - NuMI) have a smaller or equal
neutrino energy over baseline,
E/L, than the experiment with smaller matter effect (shorter baseline - JHF).
If the JHF experiment runs first
then NuMI should also run at a similar or smaller E/L to provide the 
best sensitivity
to the different mass hierarchies.
The separation between the different mass hierarchies is reduced if the 
NuMI E/L is higher than the JHF E/L.
The size of the small mixing angle,  $\theta_{13}$, cannot be determined
to better than about 30\% with this comparison alone.

The one neutrino - one anti-neutrino comparison is in general
most useful for determining
the size of the small mixing angle, $\theta_{13}$ and 
the CP or T violating phase $\delta$. 
It could also be useful in determining the mass hierarchy provided 
nature does not choose the large overlap region in the bi-probability
plot. For this comparison, the uncertainty in  $\theta_{13}$ from the
degeneracy issue is not important until the experimental resolution
in $\theta_{13}$ is better than 10\%.

\acknowledgments

HM and HN thank the Theoretical Physics Department of Fermilab for warm
hospitality extended to them during their visits.
This work was supported by the Grant-in-Aid for Scientific Research
in Priority Areas No. 12047222, Japan Ministry
of Education, Culture, Sports, Science, and Technology.
Fermilab is operated by URA under DOE contract No.~DE-AC02-76CH03000.

\section*{Appendix A}

\def \J{\alpha}
\def \N{\beta}

In this Appendix, we first explain briefly why the universal 
formalism with generic notations given in Table 1 is possible. 
The first secret behind it is the relationship between 
$X$'s and $Y$'s with different sign of $\Delta m^2_{13}$.
Let us think of the expressions of the appearance probabilities 
in neutrino and anti-neutrino channel similar to (\ref{basicequations}), 
and define coefficient function $\bar{X}_{\pm}$ as well as $X_{\pm}$. 
They can be expressed by using a single function $X$ as 
$X_{\pm} \equiv X(\pm \Delta m^2, a)$ and 
$\bar{X}_{\pm} = X(\pm \Delta m^2, -a)$.
The similar notation can also be defined for $Y$'s. 
Then, 
\begin{eqnarray}
X_{\pm} &=& \bar{X}_{\mp} 
\nonumber \\
Y_{\pm} &=& - \bar{Y}_{\mp},
\end{eqnarray}
because $X$'s and $Y$'s (except for extra sign) are the 
function only of  $\Delta_{13} \pm aL$, which follows from the 
CP-CP relation in \cite{MNP1} and the approximation introduced 
in \cite{golden}.
We note that $X_\pm$ and $Y_\pm$ satisfy the useful identity
\begin{equation}
{ Y_+ \over \sqrt{X_+}}
= - ~{ Y_- \over \sqrt{X_-}}.
\end{equation}

Next we present general solutions of (\ref{basicequations}) 
whose validity extends any combinations of channels 
tabulated in Table 1. 
For any given process and sign of $\Delta m^2_{13}$ there are 
two solutions to these equations which we will label by 
$(\theta_1, \delta_1)$ and $(\theta_2, \delta_2)$.
The equations (\ref{basicsolution}) can be solved for 
the CP phase $\delta$ for either sign of $\Delta m^2_{13}$ as: 
\begin{eqnarray}
\sin{\delta_{1,2}} &=& 
\left[
-S \Delta P  \pm 
C \sqrt{S^2+C^2 - (\Delta P)^2}
\right] /(S^2+C^2)  \nonumber \\
\cos{\delta_{1,2}} &=& 
\left[
-C \Delta P \mp 
S \sqrt{S^2+C^2 - (\Delta P)^2}
\right] /(S^2+C^2)
\label{deltas}
\end{eqnarray}
Notice that with this choice of signs 
$\sin^2 \delta_i + \cos^2 \delta_i =1$, and 
the variables $\Delta P$,  $S$, and $C$  are defined as 
\begin{eqnarray}
\Delta P &\equiv& 
\sqrt{\frac{P^\J - P^\J_{\odot}}{X^\J}} - 
\sqrt{\frac{P^\N - P^\N_{\odot}}{X^\N}}
\nonumber \\
S 
& \equiv & \left[R^\J \sin {\left(\frac{\Delta^\J}{2}\right)} 
- R^\N \sin {\left(\frac{\Delta^\N}{2}\right)} \right] 
\nonumber \\[0.1in] 
C 
& \equiv & \left[-R^\J \cos {\left(\frac{\Delta^\J}{2}\right)} 
+ R^\N \cos {\left(\frac{\Delta^\N}{2}\right)} \right], 
\end{eqnarray}
where we have defined the variable
\begin{eqnarray}
R^\J & \equiv & \frac{Y^\J}{2X^\J}
\end{eqnarray}
which appears frequently. The
corresponding $\theta$'s are then given by
\begin{eqnarray}
\theta_{1,2}  & = &
\left(
R^\J\sin{(\frac{\Delta^\J}{2})}
\sqrt{\frac{(P^\N - P^\N_{\odot})}{X^\N}}
-R^\N\sin{(\frac{\Delta^\N}{2})}
\sqrt{\frac{(P^\J - P^\J_{\odot})}{X^\J}}
\right.
\nonumber \\ 
& & + \left. 
R^\J\sin{(\frac{\Delta^\J}{2})}
R^\N\sin{(\frac{\Delta^\N}{2})}
\left(\cot{\frac{\Delta^\J}{2}}
-\cot{\frac{\Delta^\N}{2}}\right)
\cos{\delta_{1,2}} \right)
\nonumber \\
& & /
\left( {R^\J\sin{(\frac{\Delta^\J}{2})}
-R^\N\sin{(\frac{\Delta^\N}{2})}} \right).
\label{thetas} 
\end{eqnarray}
Under the interchange of $\J$ and $\N$, $\delta_1 \leftrightarrow \delta_2$
and $\theta_1 \leftrightarrow \theta_2$ as it should.
Also for useful experiments 
$\cot{\frac{\Delta^\J}{2}}$ is always finite.
These two solutions coincide, $\theta_1=\theta_2$, when the 
coefficient in front of the $\cos \delta_{1,2}$ term vanishes or
$\cos \delta_{1} = \cos \delta_{2}$.  
The first possibility occurs when $\Delta^\J = \Delta^\N$ or
$\Delta^\J = \pm \pi = \pm \Delta^\N$. 
While the second occurs at the edge of the allowed region when the
square root in Eq.(\ref{deltas}) vanishes.

The allowed region in $(P^\N,P^\J)$ bi-probability space is given by
\begin{eqnarray}
(\Delta P)^2 & \leq & S^2 + C^2 
= {(R^{\J})^2 + (R^{\N})^2 - 2 R^{\J}R^{\N}
\cos{\left(\frac{\Delta^{\J}}{2} - \frac{\Delta^{\N}}{2}\right)}}.
\label{reg_allwd}
\end{eqnarray}
This region is determined by requiring the $\sin \delta$ and $\cos \delta$
to be real.
The boundary is determined by the equality in eqn (\ref{reg_allwd}).
Thus the quantity
\begin{eqnarray}
(R^{\J})^2 + (R^{\N})^2 & - & 2 R^{\J}R^{\N}
\cos{\left(\frac{\Delta^{\J}}{2} - \frac{\Delta^{\N}}{2}\right)}.
\label{size_allwd}
\end{eqnarray}
controls the separation of the two boundaries and hence the size of the
allowed region.  Depending on the value of $(\Delta^\J - \Delta^\N)$ this can
range from $(\Delta P)^2=(R^\J-R^\N)^2$ 
when $(\Delta^\J - \Delta^\N)=0,~2\pi \cdots$ to 
$(\Delta P)^2=(R^\J+R^\N)^2$ for 
$(\Delta^\J - \Delta^\N)= \pm \pi, \pm 3\pi \cdots$.  
Which of these two extremes gives the larger
allowed region depends on the relative signs of $R^\J$ and $R^\N$.
This will be very important since the relative signs
changes as we go from both neutrino (or both anti-neutrino) to one 
neutrino and one anti-neutrino process.

Let us now compute the difference between two solutions of $\delta$. 
One can show by using (\ref{deltas}) that 
\begin{eqnarray}
\cos{(\delta_{1} + \delta_{2})} & = &
1 - 2
\frac {
\left[
R^{\J} \sin {\left(\frac{\Delta^\J}{2}\right)} - 
R^{\N} \sin {\left(\frac{\Delta^\N}{2}\right)}
\right]^2}
{(R^{\J})^2 + (R^{\N})^2 - 2 R^{\J}R^{\N}
\cos{
\left(
\frac{\Delta^{\J}}{2} - 
\frac{\Delta^{\N}}{2}
\right)}
}.
\label{delta_relation}
\end{eqnarray}
One can easily check that Eq.~(\ref{delta_relation}) reduces to 
Eq.~(45) of \cite{MNP2} if we take $\Delta^\J = -\Delta^\N$.
The relationship (\ref{delta_relation}) implies that
\begin{eqnarray}
\delta_{2} &=& \pi - \delta_{1} + 2 \arcsin{
\frac {
|R^{\J} \sin {\left(\frac{\Delta^\J}{2}\right)} - 
R^{\N} \sin {\left(\frac{\Delta^\N}{2}\right)}|}
{\sqrt{(R^{\J})^2 + (R^{\N})^2 - 2 R^{\J}R^{\N}
\cos{\left(\frac{\Delta^{\J}}{2} - \frac{\Delta^{\N}}{2}\right)}}
}
}
\label{deltadiff}
\end{eqnarray}
where use has been made of the relation 
$\arccos{(1-2 x^2)} = 2 \arcsin{x}$.
Thus the second solution $\delta_2$
differs from $\pi -\delta_1$ by a constant
 which depends on the energy and path length of 
the neutrino beams as well as on the process 
but not on the mixing angle $\theta_{13}$ for 
any two-experiment comparison. 
This generalizes the result obtained in \cite{MNP2}.

At the oscillation maximum, 
$|\Delta^{\J}|=|\Delta^{\N}| = \pi$ and 
$\cos{(\delta_{1} + \delta_{2})} = -1$ i.e
$\delta_2 = \pi -\delta_1$.
This is a reflection on the fact that at oscillation maximum the trajectory
in the bi-probability space is an ellipse with zero width as can easily
be seen from Eq.~(\ref{basicequations}) as there is no 
dependence on $\cos \delta$ at this point.

\section*{Appendix B}

We evaluate the correction to the relationship between two 
degenerate solutions of $\delta$ in $\nu$-$\nu$ and 
$\nu$-$\bar{\nu}$ comparison. 
Let us first discuss $\nu$-$\bar{\nu}$ comparison first.

We expand $S^\prime_{\pm}$, $C^\prime_{\pm}$, and 
$\Delta P^\prime_{\pm}$ at around the JHF parameters.
For simplicity, we only deal with the case $\Delta^N = \Delta^J$. 
They read,
\begin{eqnarray}
S^\prime_{\pm} &=& S^J_{\pm} \pm 
\frac{\epsilon_A}{4} 
\left(\frac{Y^J_{\mp}}{X^J_{\mp}}\right)
\sin {\left(\frac{\Delta_{13}^J}{2}\right)}
\left[
g \left(\frac{aL^J}{2} \right) \mp 
g \left(\frac{B_{\pm}}{2} \right) 
\right] \\
C^\prime_{\pm} &=& C^J_{\pm} + 
\frac{\epsilon_A}{4} 
\left(\frac{Y^J_{\mp}}{X^J_{\mp}} \right)
\cos {\left(\frac{\Delta_{13}^J}{2}\right)}
\left[
g \left(\frac{aL^J}{2} \right) \mp 
g \left(\frac{B_{\pm}}{2} \right) 
\right] \\
\Delta P^\prime_{\pm} &=& 
\Delta P^J_{\pm} \mp 
\frac{\epsilon_A}{2}
\sqrt{\frac{P^N - P^N_{\odot}}{X^J_{\mp}}}
g \left(\frac{B_{\pm}}{2} \right), 
\label{expansion}
\end{eqnarray}
where $\epsilon_A \equiv (aL)^N - (aL)^J$.
We have defined the function $g$ as 
\begin{eqnarray}
g(x) \equiv \frac{1}{x} - \cot{x}.
\end{eqnarray}
The function $g$ monotonically increases in 
$0 \leq x \leq \pi$.

Then, $\sin{(\delta_{1} - \delta_{3})}$ in Eq.~(\ref{del-diffs}) 
can be expanded to first order in $\epsilon_A$ as 
$\sin{(\delta_{1} - \delta_{3})} = C^{N-J} \epsilon_A$. 
The coefficient $C^{N-J}$ is given by 
\begin{eqnarray}
C^{N-J} &=&  
\frac{S^J C^J}{D^J} g \left(\frac{aL^J}{2} \right) +  
\frac{1}{4}
\frac{\Delta P_0}{\sqrt{D^J - \Delta P_0^2}} 
\left[
g \left(\frac{B^J_{+}}{2} \right) + 
g \left(\frac{B^J_{-}}{2} \right)
\right] 
\nonumber \\
&+&
\frac{1}{2 \sin{\Delta^J}} \frac{S^J C^J}{D^J} 
\frac{\Delta P_0}{\sqrt{D^J - \Delta P_0^2}} 
\left[
g \left(\frac{B^J_{+}}{2} \right) - 
g \left(\frac{B^J_{-}}{2} \right) - 
2 g \left(\frac{aL^J}{2} \right) 
\right] 
\nonumber \\
&-& 
\frac{1}{8}
\frac{\sin{\Delta^J}}{D^J} 
\left[
\left(\frac{Y^J_{+}}{X^J_{+}} \right)^2
g \left(\frac{B^J_{-}}{2} \right) + 
\left(\frac{Y^J_{-}}{X^J_{-}} \right)^2
g \left(\frac{B^J_{+}}{2} \right)
\right] 
\nonumber \\
&-& 
\frac{1}{2}
\frac{1}{\sqrt{D^J - \Delta P_0^2}} 
\left[
\sqrt{\frac{P^N - P^N_{\odot}}{X^J_{+}}}
g \left(\frac{B^J_{-}}{2} \right) - 
\sqrt{\frac{P^N - P^N_{\odot}}{X^J_{-}}}
g \left(\frac{B^J_{+}}{2} \right)
\right]
\label{C^{N-J}}
\end{eqnarray}
where 
$S^J$, $C^J$, and $\Delta P_0$ indicate, respectively, 
the $\epsilon_A \rightarrow 0$ limit of 
$S^\prime_{\pm}$, $C^\prime_{\pm}$, and 
$\Delta P^\prime_{+}$ (or superscript $N$ replaced by $J$). 
Notice that 
$\Delta P^\prime_{+}= - \Delta P^\prime_{-} \equiv \Delta P_0$ 
in the limit $P^N=P^J$ and that 
$S^J_{+}=S^J_{-}=S^J$ and$C^J_{+}=C^J_{-}=C^J$.

Now we turn to $\nu$-$\nu$ comparison. In this case there is no 
zeroth order term in $S_{\pm}$ etc. for reasons explained 
in subsection IIA. Instead they start with the first-order 
terms in $\epsilon_A$. We do not give the details but just 
mention that $\sin{(\delta_{1} - \delta_{3})}$ is 
given to first-order by 
\begin{eqnarray}
\sin{(\delta_{1} - \delta_{3})} &=&  
\frac{P^J}
{
\frac{Y^J_{+}Y^J_{-}}{\sqrt{X^J_{+}X^J_{-}}} 
\left[
1 + \frac{g \left(\frac{aL^J}{2} \right)}{g \left(\frac{B^J_{-}}{2} \right)}
\right] 
\left[
1 - \frac{g \left(\frac{aL^J}{2} \right)}{g \left(\frac{B^J_{+}}{2} \right)}
\right]
} 
\nonumber \\
&\times&
\left[
\sin{(\Delta^J)}
\left(1 + 
\sqrt{
\frac{(Y^J_{+})^2}{4 X^J_{+} P^J} 
\left[
1 + \frac{g \left(\frac{aL^J}{2} \right)}{g \left(\frac{B^J_{-}}{2} \right)}
\right]^2 - 1}
\sqrt{
\frac{(Y^J_{-})^2}{4 X^J_{-} P^J} 
\left[
1 + \frac{g \left(\frac{aL^J}{2} \right)}{g \left(\frac{B^J_{+}}{2} \right)}
\right]^2 - 1}
\right) \right.
\nonumber \\
&-& \left.
\cos{(\Delta^J)}
\left(
\sqrt{
\frac{(Y^J_{+})^2}{4 X^J_{+} P^J} 
\left[
1 + \frac{g \left(\frac{aL^J}{2} \right)}{g \left(\frac{B^J_{-}}{2} \right)}
\right]^2 - 1} 
- 
\sqrt{
\frac{(Y^J_{-})^2}{4 X^J_{-} P^J} 
\left[
1 + \frac{g \left(\frac{aL^J}{2} \right)}{g \left(\frac{B^J_{+}}{2} \right)}
\right]^2 - 1}
~\right)
\right]
\end{eqnarray}
One can show that ($\delta_{1} - \delta_{3}$) is in the second quadrant 
because $\cos{(\delta_{1} - \delta_{3})} < 0$ in our convention.


\begin{figure}
\vspace*{16.3cm}
\includegraphics{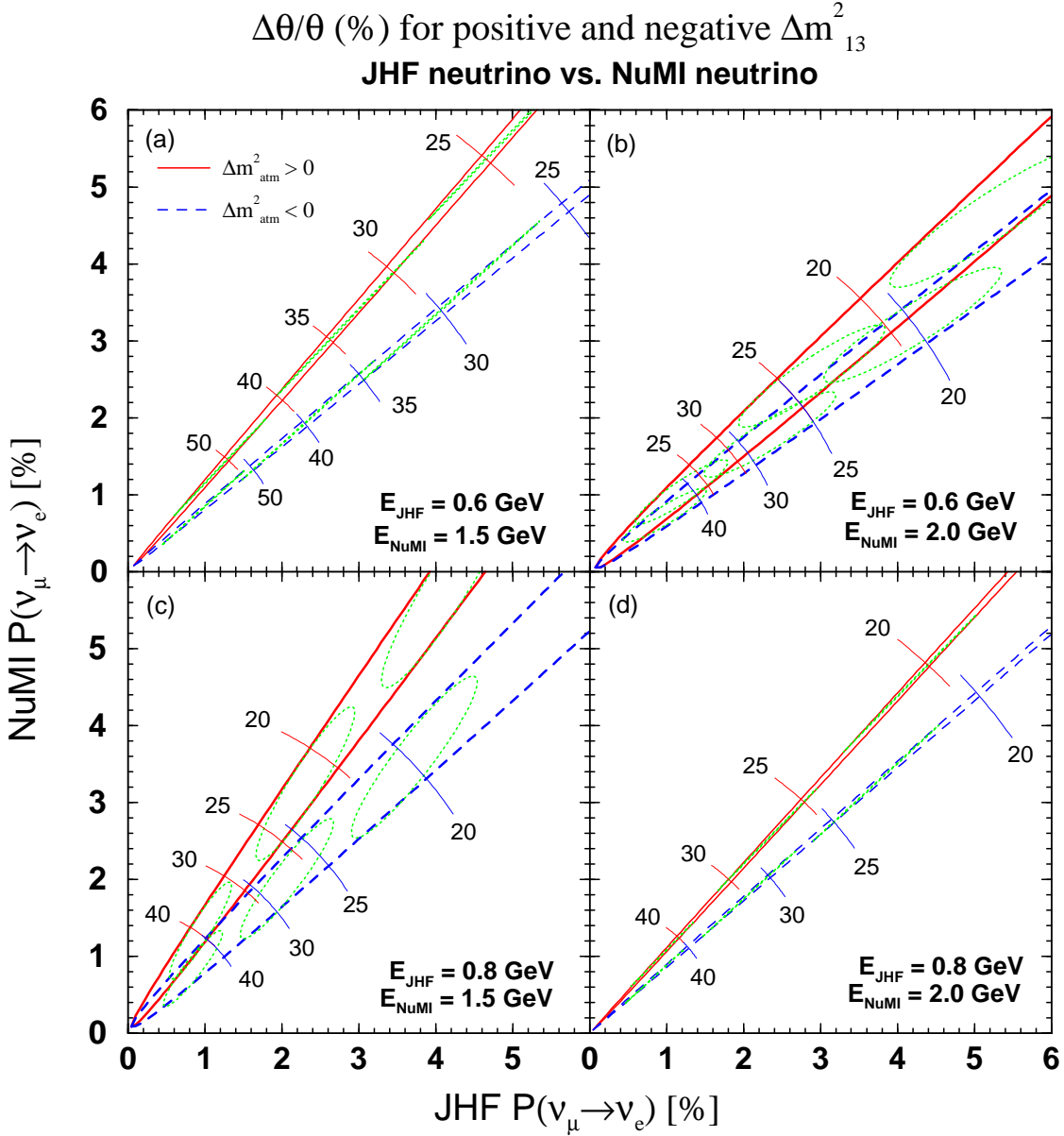}
\vspace{0.5cm}
\caption[]{
Allowed range of $P(\nu_\mu \rightarrow \nu_e)$ for JHF
verses $P(\nu_\mu \rightarrow \nu_e)$ for NuMI,
which are referred to as ``pencils'' in the text, 
are delimited by thick solid (dashed) lines for positive (negative)
$\Delta m^2_{13}$ for the energies ($E_{JHF}$/GeV, $E_{NuMI}$/GeV) 
= (a) (0.6,1.5), (oscillation maximum for both experiments) 
(b) (0.6,2.0), (c) (0.8,1.5) 
and (d) (0.8,2.0). 
In the same plot, 
the positions for some representative values of 
the fractional 
variation across the width of the ``pencil'' of
$\theta \equiv \sin \theta_{13}$, 
$\Delta \theta/\theta$ [\%], indicated by numbers, 
are shown by thin solid arcs.  
Inside each allowed region, trajectories corresponding to 
$\sin^2 2\theta_{13}=0.02$, 0.05 and 0.09 are plotted
by dotted lines.
The mixing parameters are fixed to be 
$|\Delta m^2_{13}| = 2.5 \times 10^{-3}$ eV$^2$,
$\sin^2 2\theta_{23}=1.0$,
$\Delta m^2_{12} = +7 \times 10^{-5}$ eV$^2$
and $\sin^2 2\theta_{12}=0.85$
whereas $\theta_{13}$ and $\delta$ are assumed to be unknown. 
The electron density is fixed to be 
$ Y_e \rho  = 1.15$ and  1.4g cm$^{-3}$ for 
JHF and NuMI experiment, respectively.
For JHF and NuMI both anti-neutrinos, the roles of 
$\Delta m^2_{13} > ~0 ~{\rm and} ~\Delta m^2_{13} < ~0$ are interchanged.
}
\label{nunufig}
\end{figure}

\begin{figure}
\vspace*{17cm}
\includegraphics{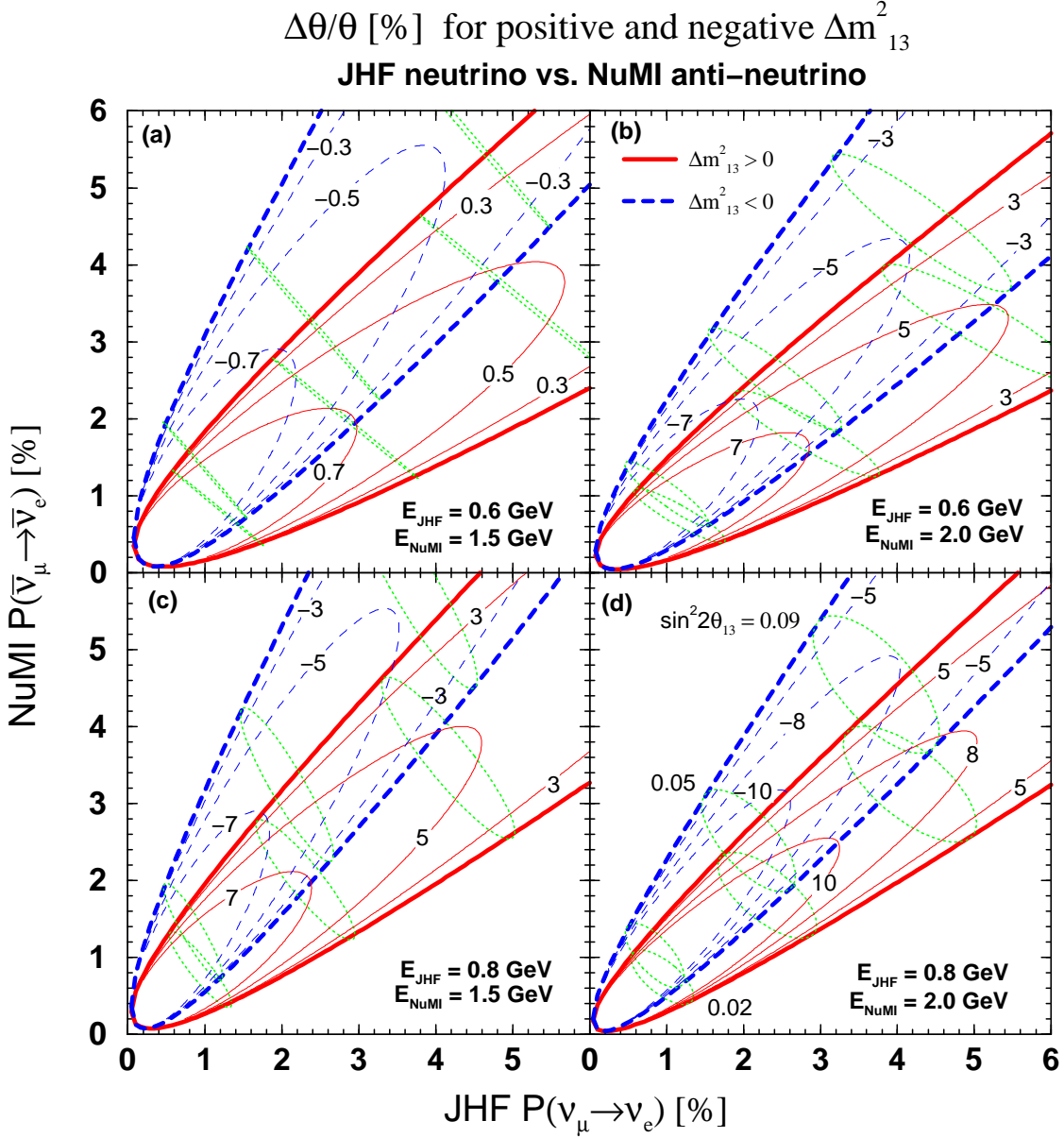}
\vspace{1.7cm}
\caption[]{
Iso-contours of $\Delta \theta/\theta$ [\%] are indicated in
the allowed range of $P(\nu_\mu \rightarrow \nu_e)$ for JHF
verses $P(\bar{\nu}_\mu \rightarrow \bar{\nu}_e)$ for NuMI
delimited by solid (dashed) lines for positive (negative)
$\Delta m^2$ for 
the same combinations of energies as in Fig.~\ref{nunufig}.  
Inside each allowed region, trajectories corresponding to 
$\sin^2 2\theta_{13}=0.02$, 0.05 and 0.09 are plotted
by dotted lines.
The assumption about the mixing parameters are 
the same as in Fig.~\ref{nunufig}.
For JHF anti-neutrinos and NuMI neutrinos, the roles of
$\Delta m^2_{13} > ~0 ~{\rm and} ~\Delta m^2_{13} < ~0$ are interchanged.
}
\label{nuanufig1}
\end{figure}

\begin{figure}
\vspace*{17cm}
\includegraphics{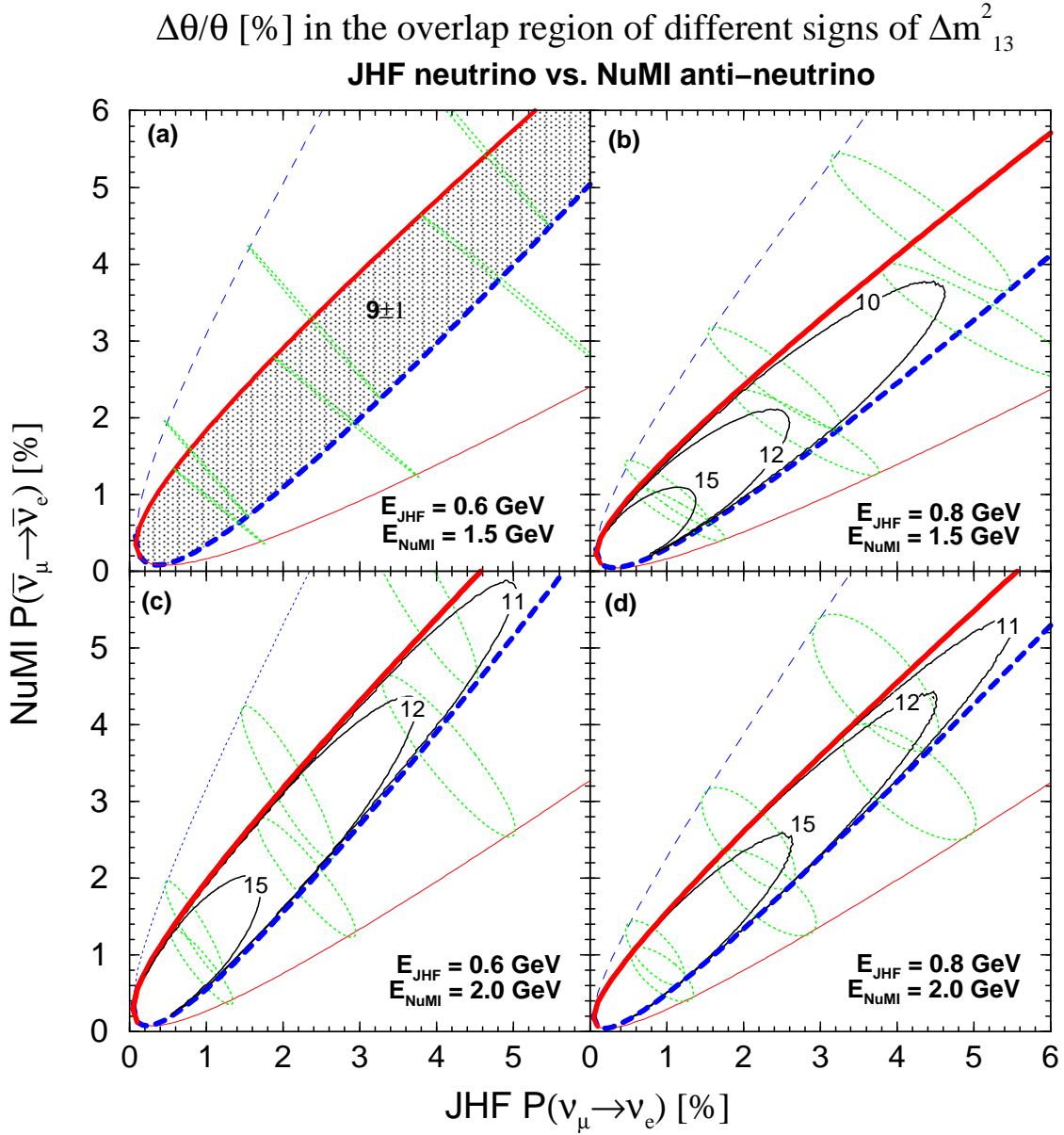}
\vspace{1.7cm}
\caption[]{
Iso-contours of 
$\Delta \theta/\theta$ [\%], 
except for (a), are indicated in
the overlap regions of the allowed range of 
$P(\nu_\mu \rightarrow \nu_e)$ for JHF
verses $P(\bar{\nu}_\mu \rightarrow \bar{\nu}_e)$ for NuMI
for the positive and negative signs of $\Delta m^2_{13}$ 
for the combinations of energies shown in 
Fig.~\ref{nuanufig1}. For (a), inside the shaded region,
$\Delta \theta/\theta$ varies very little, taking values 
close to 9 \%. 
}
\label{nuanufig2}
\end{figure}

\end{document}